\documentclass{amsart}

\usepackage{url}
\usepackage{multirow}
\usepackage{tikz}
\usetikzlibrary{arrows,positioning,shapes.geometric}
\usepackage{wrapfig}

\newtheorem{theorem}{Theorem}[section]

\theoremstyle{definition}

\theoremstyle{remark}
\newtheorem{remark}[theorem]{Remark}

\numberwithin{equation}{section}

\newcommand{\ours}{\texttt{SAVER}}

\author[R.~Li]{Ruilin Li}
\address{Department of Mathematics, University of Toronto, Toronto, Ontario, Canada}
\email{ruilin.li@mail.utoronto.ca}

\author[F.~Martin]{Martin G. Frasch}
\address{Department of Obstetrics and Gynecology, University of Washington, Seattle, WA, USA}
\email{mfrasch@uw.edu}

\author[H.-T.~Wu]{Hau-Tieng~Wu}
\address{Department of Mathematics, University of Toronto}
\email{hauwu@math.toronto.edu}

\title[Two channel fECG via diffusion]{Efficient fetal-maternal ECG signal separation from two channel maternal abdominal ECG via diffusion-based channel selection}

\begin{document}

\begin{abstract}

There is a need for affordable, widely deployable maternal-fetal ECG monitors to improve maternal and fetal health during pregnancy and delivery. Based on the diffusion-based channel selection, here we present the mathematical formalism and clinical validation of an algorithm capable of accurate separation of maternal and fetal ECG from a two channel signal acquired over maternal abdomen. 
\end{abstract}

\maketitle

\section{Introduction}

Fetal electrocardiogram (ECG) and the fetal heart rate (HR) provide enormous information about fetal health. For example, the fetal distress monitoring \cite{Jenkins1989} or the potential risk for fetal hypoxia detection and alert by the ST analysis monitor \cite{Belfort_Saade:2015}.
Moreover, from clinical studies and animal models, evidence is accumulating that perinatal brain injury originates in utero, yet no means exist to detect its onset early, reliably and with simple, widely accessible means \cite{Anblagan:2016}. A harbinger of brain injury is the fetal inflammatory response \cite{Hagberg:2015}. There is an urgent need for early antenatal detection of fetal inflammatory response to prevent or at least mitigate the developing perinatal brain injury. 
In adults and neonates, complex mathematical features of heart rate fluctuations have proven promising as early diagnostic tools \cite{Bravi:2013,Fairchild:2014}. For the fetal monitoring, our team addressed the challenge by developing a series of biomarkers relying on non-invasively obtainable fetal HR. Our fetal inflammatory index tracks inflammation along with the fetal plasma IL-6 temporal profile in a fetal sheep model of subclinical chorioamnionitis \cite{Durosier:2015}. We also derived a set of fetal HR features that is specific to brain or gut inflammation \cite{Liu:2016}. Such systemic and organ-specific tracking of inflammation via fetal HR is possible due to the brain-innate immune system communication reflected in the fetal HR fluctuations, commonly referred to as the cholinergic anti-inflammatory pathway \cite{Garzoni:2013,Olofsson:2012,Fairchild:2011}.

In spite of its broad usefulness in the fetal health, it is fair to state that in the fetal HR monitoring realm, the technological progress has been coming more gradually. This has been not due to the plethora of studies attempting and testing various approaches, but, rather, due to the intrinsic limitations of the currently used fetal HR monitoring technology. This technology is outdated, as it deploys the traditionally set low sampling rate of heart rate or ECG signal. In animal model and human cohorts, we showed that such sampling rate is bound to miss the faster temporal fluctuations of vagal modulations of fetal HR variability and leads to inaccuracies in detection of early fetal acidemia \cite{Durosier:2014,Li:2015}. A sampling rate of the ECG signal around 1000 Hz is required to capture these vagal influences and this is the commonly used sampling rate for the postnatal studies and our above-cited studies on the fetal inflammatory index.

Postnatal clinical studies are typically based on multi-lead ECG recordings which, even in newborns, and certainly in adults, poses no technical challenge to attach and record from. In fetuses, however, this is not the case. Since the fetal cardiac electric field strength is order of magnitude weaker than maternal ECG's, and the lack of clinical motivation in higher quality fetal HR data, little development had been done to focus on fetal ECG (fECG) signal in the clinical monitoring until today, except the Doppler-based fetal HR extraction techniques that dominate the market. The Doppler-based fetal HR extraction techniques, however, suffer from low fetal HR sampling rates, largely due to the auto-correlation algorithms deployed in the devices \cite{Durosier:2014}. Transabdominal ECG (aECG) machines overcome this limitation by capturing the actual cardiac electric field and have returned to the market during the last decade. However, their arrival has been slower than we would have hoped. Perhaps this is in part due to the general acceptance speed of new technology in medicine (related to regulatory and safety testing as well as the specific cultures), due to the high cost for each device to upgrade a hospital's delivery unit, or, more likely, the technical limitation of the fetal ECG extraction from the aECG signals. To make the technology of high quality and low-cost fetal ECG widely accessible, we need algorithms for fetal ECG extraction from easily deployable aECG devices. 

The current study addresses this challenge by proposing an algorithm capable of working with only two composite (maternal and fetal) aECG channels to derive the fetal signal from it.
It is based on the currently developed single-lead fECG algorithm based on the modern time-frequency analysis and manifold learning technique \cite{Su_Wu:2016} and a novel proposed diffusion-based channel selection criteria. All the proposed methods have rigorous mathematical backups, and numerically they can be efficiently implemented to handle long signal. We call the proposed algorithm $\ours$, which stands for \textbf{S}mart \textbf{A}dapti\textbf{V}e \textbf{E}cg \textbf{R}ecognition. To validate $\ours$, we report the analysis results of two publicly available databases, and compare the algorithm with other available algorithms in the literature. 

The paper is organized in the following way. 
In Section \ref{Section:ProposedAlgorithm}, we detail our proposed algorithm, describe the algorithms we will compare, and describe the databases we validate the algorithm. The results are shown in Section \ref{Section:Results}, and the discussions with the future works are provided in Section \ref{Section:Discussions}. The paper closes with the conclusion shown in Section \ref{Section:Conclusion}
The necessary theoretical background is provided in $\ours$ Section \ref{Section:TheoreticalBackground}, particularly the diffusion-based channel selection criteria. We refer the readers to \cite{Su_Wu:2016} for the details of the de-shape short time Fourier transform (dsSTFT), beat tracking and the nonlocal median.

\section{Methods}\label{Section:ProposedAlgorithm}

\subsection{Two-lead fECG Algorithm -- $\ours$}

We now describe the proposed two-channel fECG algorithm, which the authors coined as $\ours$. 
The overall algorithm is illustrated in Figure \ref{FlowChart}.

\tikzstyle{line} = [draw, -latex']
\tikzstyle{arrow} = [thick,->,>=stealth]

\begin{figure}[h!]
\centering
   \begin{tikzpicture}[>=latex']
        \tikzset{
        block/.style= {draw, rectangle, align=center,minimum width=12cm,minimum height=.10cm,line width=0.3mm},
        rblock/.style={draw, shape=rectangle,rounded corners=1.5em,align=center,minimum width=12cm,minimum height=.10cm},
        }

        \node [rblock]  (start) {
        \includegraphics[width=.7\textwidth]{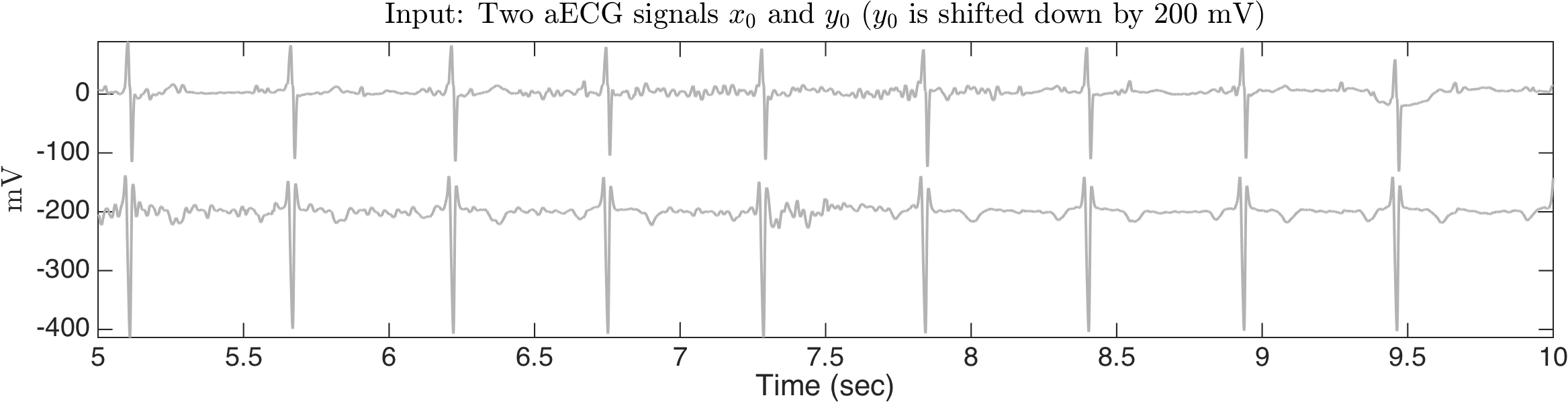}};
        \node [block, below =.5cm of start] (Step1) {\includegraphics[width=.7\textwidth]{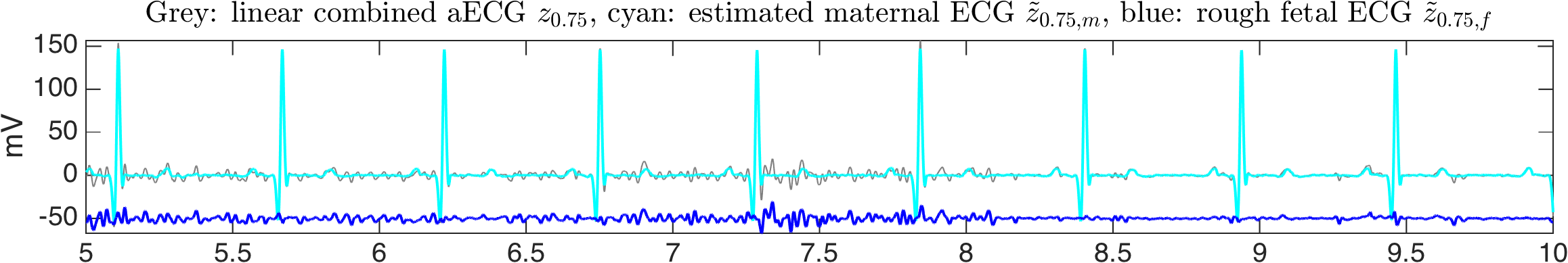}\\\includegraphics[width=.7\textwidth]{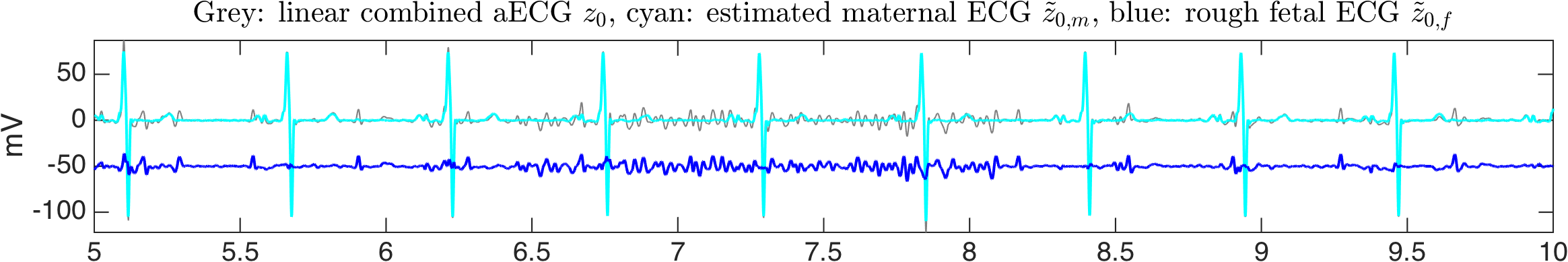}\\\includegraphics[width=.7\textwidth]{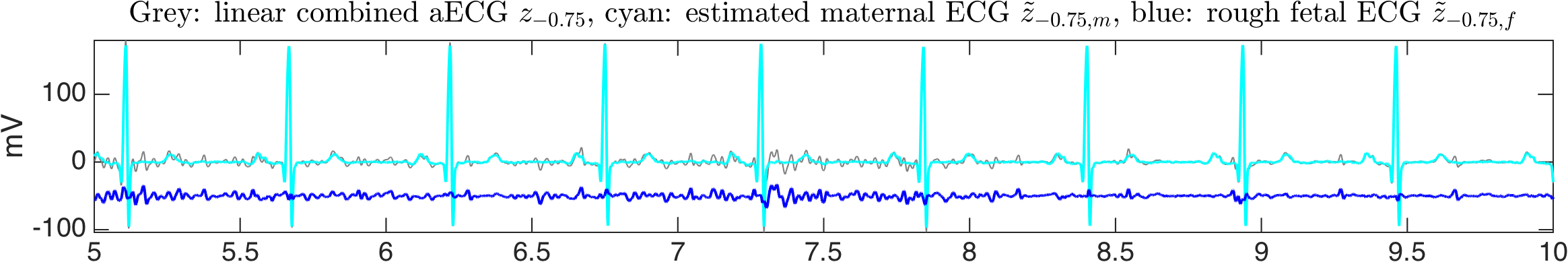}};
        \node [block, below =.5cm of Step1] (Step2) {\includegraphics[width=.7\textwidth]{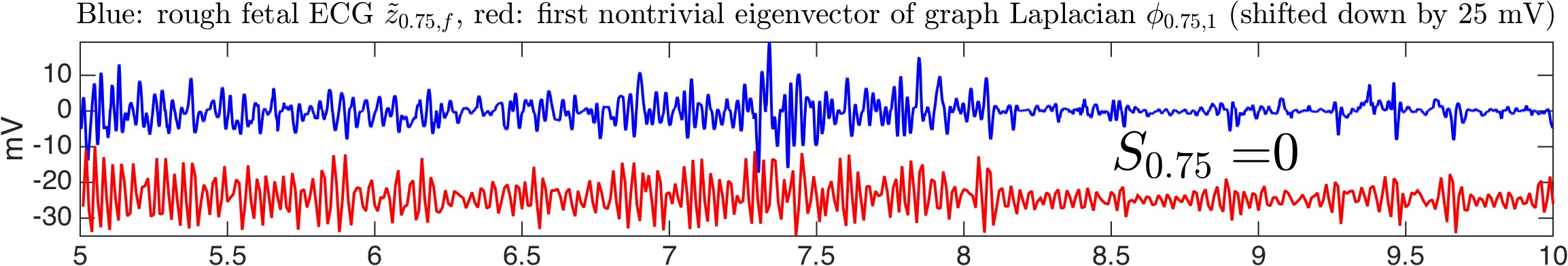}\\\includegraphics[width=.7\textwidth]{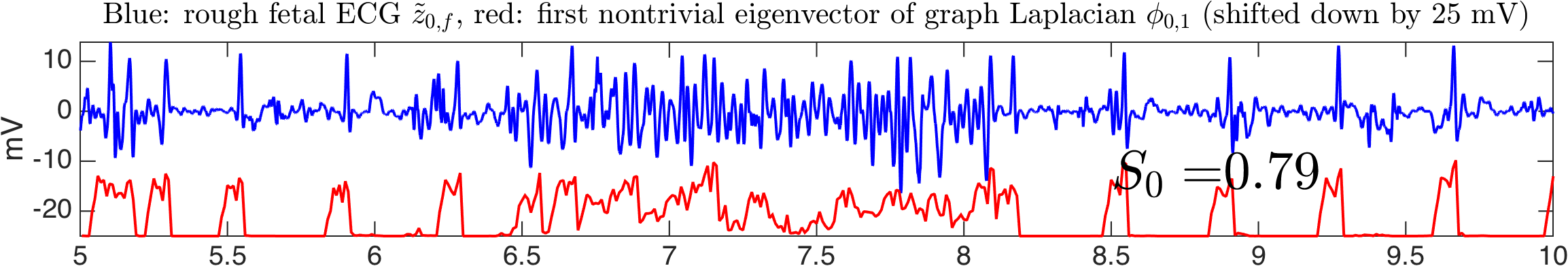}\\\includegraphics[width=.7\textwidth]{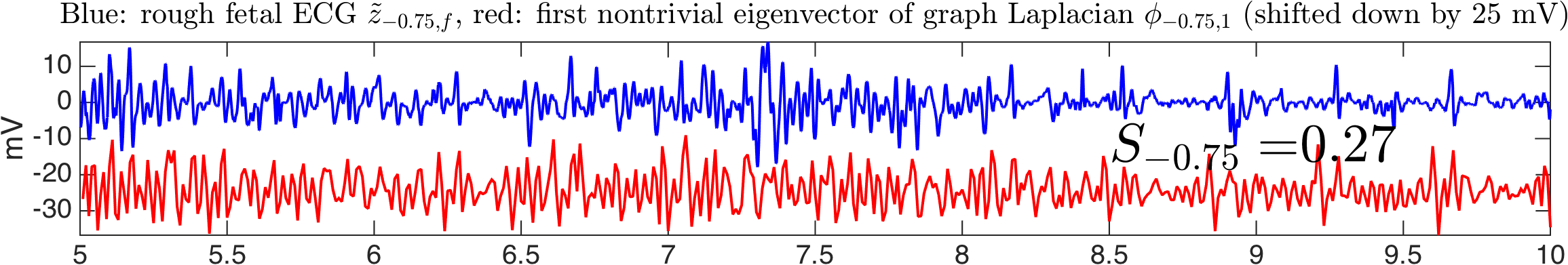}};
        \node [rblock, below =.5cm of Step2] (Final) {\includegraphics[width=.7\textwidth]{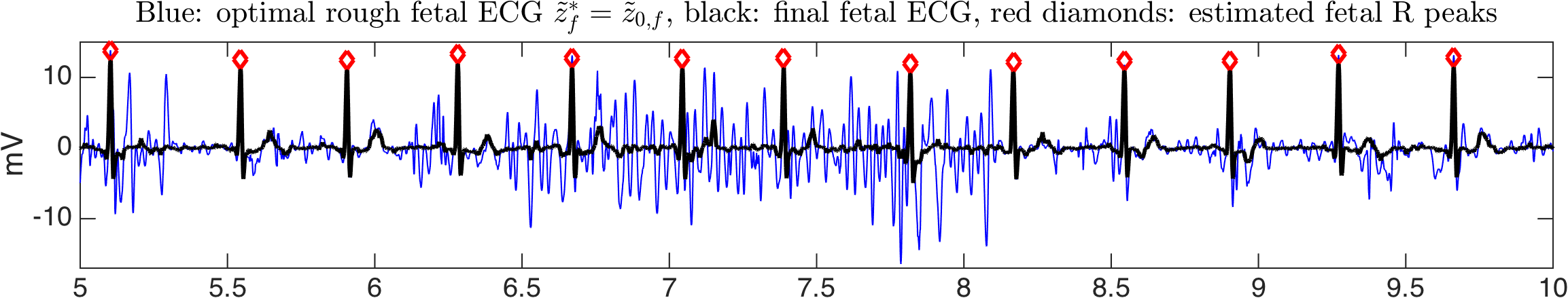}};

          \draw[arrow]   (start.south)  --(Step1.north) node [pos=0.66,right] {Step 0 + Step 1};
          \draw[arrow]   (Step1.south)  --(Step2.north) node [pos=0.66,right] {Step 2};
          \draw[arrow]   (Step2.south)  -- (Final.north) node [pos=0.66,right] {Step 3};
    \end{tikzpicture}
    \caption{\label{FlowChart}The flow chart of the proposed two-channel fECG algorithm, $\ours$. The x-axis of all figures are of the unit second. The data is the a2 recording from the database used in the 2013 PhysioNet/Computing in Cardiology Challenge, and channel 1 and channel 4 are shown in this illustration. Only three linear combinations are shown for the illustration purpose. The signal quality index for the channel selection is shown on the third block.}
\end{figure}

Denote two simultaneously recorded aECG signals as $\mathbf{x}_0, \mathbf{y}_0\in \mathbb{R}^N$ with the sampling rate $\xi_0$Hz over the interval from the $0$-th second to the $N/\xi_0$-th second. If the signal is sampled more slowly than 1000 Hz, to enhance the R peak detection and the the nonlocal median \cite{Su_Wu:2016}, the signal is upsampled to 1000 Hz \cite{Laguna_Sornmo:2000}. We use the same notations to denote the upsampled signal.

\textbf{Step 0: pre-processing.} To suppress the noise, the signal is low-pass filtered below 100 Hz. Then, subtract the estimated trend from $\mathbf{x}_0, \mathbf{y}_0$, where the trends are estimated using median filter with window length $L_{\mathtt{MF}}>0$ second. 
If needed, the power-line interference is suppressed by two notch-filters at $50$ Hz and $60$ Hz, since the origin of the tested database in this paper is unknown (if the resource of the database is known, the notch-filter will be designed according to the power system of that region). 
Denote the pre-processed signal as $\mathbf{x}$ and $\mathbf{y}$. 
Take a discrete finite subset $\mathcal{I}\subset (-1,1]$. Define $\mathbf{z}_{\theta} = \theta \mathbf{x} + \sqrt{1-\theta^2} \mathbf{y}$, where $\theta\in \mathcal{I}$; that is, $\mathbf{z}_{\theta}$ is a linear combination of two aECG signals. This linear combination could be viewed as a generalization of the augmentation technique considered in \cite[Section 2.3.3]{AndreottiRiedl2014}.

\textbf{Step 1: maternal ECG estimation}

We iterate the dsSTFT and nonlocal median algorithms proposed in \cite{Su_Wu:2016}  to decompose the maECG from each linear combination in $\{\mathbf{z}_{\theta}\}_{\theta\in \mathcal{I}}$. The algorithm is summarized below. For each $\theta$ we run the following three sub-steps.

\begin{enumerate}
    \item (step 1-1) Apply the dsSTFT to $\mathbf{z}_{\theta}$ and extract the dominant curve in the dsSTFT \cite[Section 3.1.2]{Su_Wu:2016}, which represents the estimated maternal IHR. 
    
    \item (step 1-2) Compute the polarity of $\mathbf{z}_{\theta}$, where the polarity is either positive or negative. If the polarity of $\mathbf{z}_{\theta}$ is negative, multiply $\mathbf{z}_{\theta}$ by $-1$; that is, flip the sign of $\mathbf{z}_{\theta}$. We use the same notation $\mathbf{z}_{\theta}$ to denote the polarity-corrected ECG signal. 
    With the estimated maternal IHR and the polarity-corrected ECG signal, apply the beat tracking algorithm \cite[Section 3.1.3]{Su_Wu:2016} to $\mathbf{z}_{\theta}$ to compute the locations of maternal R-peaks. Denote the timestamps of estimated maternal R peaks as $\mathbf{r}_\theta^m = (r^m_{\theta,1} , \dots, r^m_{\theta,k_{\theta,m}})$, where $k_{\theta,m}\in\mathbb{N}$ is the number of estimated maternal R peaks.

    \item (step 1-3) Adjust the estimated maternal R-peak locations by searching the maximum of $\mathbf{z}_{\theta}$ over a small window around $\mathbf{r}_\theta^m$. We use the same notation $\mathbf{r}_\theta^m$ to denote the adjusted estimated maternal R-peak locations. Apply the nonlocal median \cite[Section 3.1.4]{Su_Wu:2016} to estimate the maECG in $\mathbf{z}_{\theta}$ based on the estimated R-peak locations $\mathbf{r}_\theta^m$. Denote the estimated maECG as $\tilde{\mathbf{z}}_{\theta,m}$. 
\end{enumerate}

\textbf{Step 2: channel selection}

For each linear combination in $\{{\mathbf{z}}_{\theta}\}_{\theta\in \mathcal{I}}$, with the estimated maECG, we obtain a rough fECG by a simple subtraction: 
\begin{equation}
\tilde{\mathbf{z}}_{\theta,f}:=\mathbf{z}_{\theta}-\tilde{\mathbf{z}}_{\theta,m}.\label{Algorithm:Step2:CS}
\end{equation}
Denote $\{\tilde{\mathbf{z}}_{\theta,f}\}_{\theta\in \mathcal{I}}$ to be the set of rough fECG signals estimated from Step 1.
We apply the lag map and the diffusion map (DM) to each rough fECG in $\{\tilde{\mathbf{z}}_{\theta,f}\}_{\theta\in \mathcal{I}}$ and select the optimal linear combination by the following procedure. See Section \ref{Section:TheoreticalBackground} in the Appendix for the theoretical background of this approach.

For each rough fECG, say $\tilde{\mathbf{z}}_{\theta,f}$, we evaluate the signal quality index (SQI) for the channel selection purpose in the following way. 
Apply the $L$-step lag map to embed the interval $[2,T_{\texttt{CS}}+2]$ seconds of $\tilde{\mathbf{z}}_{\theta,f}$ into $\mathbb{R}^L$, where $T_{\texttt{CS}}>0$ is chosen by the user and $2$ is chosen to avoid the boundary effect associated with the window in the dsSTFT approach. Here $T_{\texttt{CS}}$ is chosen to be short enough to guarantee the computational efficiency and to avoid the possibility nonstationarity inherited in the fECG signal, and long enough to capture the periodicity of the fECG.
Denote the embedded point cloud as $\mathcal{X}_{\theta,f}\subset \mathbb{R}^L$. Apply the $1$-normalization DM to $\mathcal{X}_{\theta,f}$, where the bandwidth of the kernel is chosen in the following way suggested in \cite{5210209}. We first set $\epsilon_0$ to be the smallest value such that each data point has at least one neighbour within the distance $\epsilon_0$. Then we set the bandwidth to be $2\epsilon_0$. Denote $\phi_{\theta,1}$ be the first nontrivial eigenvector of the corresponding graph Laplacian.
Compute the power spectrum of $\phi_{\theta,1}$, denoted as $|\hat{\phi}_{\theta,1}|^2$.
Denote $\xi_{\theta,1},\xi_{\theta,2},\ldots,\xi_{\theta,n_{\texttt{CS}}}>0$, where $n_{\texttt{CS}}\in\mathbb{N}$ is the number of peaks chosen by the user, to be the frequencies associated with the highest $n_{\texttt{CS}}$ peaks in $|\hat{\phi}_{\theta,1}|^2$. Fix $L_{\texttt{CS}}>0$ and denote $\mathcal{J}_{\theta}:=\cup_{j=1}^{n_{\texttt{CS}}}[\xi_{\theta,i}-L_{\texttt{CS}},\,\xi_{\theta,i}+L_{\texttt{CS}}]$. The SQI for the channel selection purpose is thus defined as
\begin{equation}
S_\theta=\frac{\int_{[0,\xi_0/4)\cap\mathcal{J}_{\theta}}|\hat{\phi}_{\theta,1}(\xi)|^2d\xi}{\int_{[0,\xi_0/2)\backslash\mathcal{J}_{\theta}}|\hat{\phi}_{\theta,1}(\xi)|^2d\xi}.
\end{equation}
Under the assumption that the better the quality of the rough fECG is, the closer the embedded point cloud is to the one-dimensional circle, we know that the higher the SQI, the better the rough fECG is. More precisely, if the embedded point cloud is close to the one-dimensional circle, the first non-trivial eigenvector should behave like an oscillatory function. 
With the designed SQI, we could choose the optimal rough fECG as the one with the highest SQI.
Denote $\tilde{\mathbf{z}}^{*}_f$ to be the optimal rough fECG with the highest signal quality index we can obtain from the given two channels.

\textbf{Step 3: fetal R peaks estimation}

With the rough fECG $\tilde{\mathbf{z}}^{*}_f$ obtained from the optimal linear combination, we finish the algorithm by estimating the fetal R peaks and fECG by again applying the dsSTFT and the nonlocal median algorithm. This part of the algorithm is essentially the same as that for the maternal ECG estimation, and we repeat the three sub-steps below for the sake of completeness.

\begin{enumerate}

    \item (step 3-1) Apply the dsSTFT to $\tilde{\mathbf{z}}^{*}_f$ and extract the dominant curve in the dsSTFT, which represents the estimated fetal IHR. 
    
    \item (step 3-2) Compute the polarity of $\tilde{\mathbf{z}}^{*}_f$. If the polarity of $\tilde{\mathbf{z}}^{*}_f$ is negative, multiply $\tilde{\mathbf{z}}^{*}_f$ by $-1$, and use the same notation $\mathbf{z}^{*}_f$ to denote the polarity-corrected ECG signal. 
    With the estimated fetal IHR and the polarity-corrected ECG signal, apply the beat tracking algorithm to $\tilde{\mathbf{z}}^{*}_f$ to compute the locations of maternal R-peaks. Denote the timestamps of estimated fetal R peaks as $\mathbf{r}^f = (r^f_1 , \dots, r^f_{k_f})$, where $k_f\in\mathbb{N}$ is the number of estimated fetal R peaks.

    \item (step 3-3) Adjust the estimated fetal R-peak locations by searching the maximum of $\tilde{\mathbf{z}}^{*}_f$ over a small window around $\mathbf{r}^f$, and use the same notation $\mathbf{r}^f$ to denote the adjusted estimated fetal R-peak locations. Finally, output the fetal R peaks.

\end{enumerate}

\begin{remark}
We mention that by applying the nonlocal median again based on $\mathbf{r}^f$, we could denoise the optimal rough fECG waveform $\tilde{\mathbf{z}}^{*}_f$ and obtain a clean fetal waveform. However, since the result is similar to that shown in \cite{Su_Wu:2016}, and the focus of this paper is the fetal R peak detection, we skip the details of the fECG reconstruction in this study, and leave the fetal waveform reconstruction in the future work. 
\end{remark}

\subsection{Comparison with other algorithms}

There have been several algorithms proposed in the field suitable for analyzing fECG from multiple channel aECG signals. 
Note that the two-channel aECG signals fall in the category of the blind source separation (BSS) \cite{Lathauwer2000,Akhbari2013,DiMariaLiu2014,Varanini2014} and its variations \cite{SameniJutten2008,Haghpanahi2013,Akbari2015}. It is well known that usually we need more than 4 channels to have a reasonable result \cite{Andreotti2016}. Due to the stationarity assumption of the ICA, the input signal should be truncated to be short enough, like 30 seconds long. An important step in the BSS approach is channel selection, which is critical to identify the decomposed channel that contains the maternal or fetal ECG. Although we only have two channels, for the comparison purpose, we still show the results of the BSS approaches, including the joint approximation diagonalization of eigen-matrices (JADE) for the independent component analysis (ICA) and the principal component analysis (PCA). Since there are only two decomposed signals, we do not carry out the channel selection algorithms proposed in, for example, \cite{AndreottiRiedl2014}; instead, we take the ground truth annotation to select the optimal channel that is more likely to be the fECG, and report the detected R peaks from this detected channel. Note that we do not take the ground truth annotation into account in any other algorithms considered in this paper except this BSS approach, due to the limited number of channels. We apply the publicly available PCA and ICA codes provided in \url{http://www.fecgsyn.com}.

Another set of algorithms allow us to take only single mECG signal, but need to simultaneously acquire the maternal thoracic-lead ECG signal (tECG). Examples include
adaptive filtering (AF) based on the least mean square (AF-LMS) \cite{Widrow1975} or the recursive least square (AF-RLS) \cite{Behar2014} and its variations, like the echo state neural network (ESN) \cite{Behar2014}, blind adaptive filtering \cite{Graupe2008},
extended Kalman filter \cite{Sameni2008,Niknazar2013,AndreottiRiedl2014}, etc.
In these algorithms, the maternal thoracic ECG signal (mtECG) is needed and is viewed as the reference channel. The mtECG contains the maternal cardiac activity information that we want to remove from the aECG. Based on the assumption that the mtECG and the maternal cardiac activity in the aECG are linearly related, the AF-LMS or AF-RLS helps to extract the fECG from the aECG by removing the maternal cardiac activity in the aECG. If the relationship between the tECG and the maternal cardiac activity in the aECG is nonlinear, then ESN could help.  
However, it is not always the case that we could get the mtECG, particularly in our setup, so these algorithms could not be directly applied for our purpose. 
Since it has been shown in \cite{Su_Wu:2016} that by combining the dsSTFT and nonlocal median, we are able to estimate the maECG signal accurately. We could thus view the estimated maECG signal as the reference channel. This consideration can also be found in, for example, \cite{Rodrigues2014}. 
We thus consider the following combinations of the proposed two channel fECG algorithm and the AF-LMS or ESN. Precisely, in our proposed algorithm, we replace the direct subtraction (\ref{Algorithm:Step2:CS}) in Step 2 by the AF-LMS or ESN, by taking the estimated maECG as the reference channel to get the rough fECG. We call the combined algorithm ds-AF-LMS or ds-ESN. Note that under the assumption that the nonlocal median does a good job to recover the maECG, the reference channel should be the same as, or linearly related to, the maternal cardiac activity in the aECG, so the AF-LMS could be applied. Note that the same idea could be applied to other algorithms, like AF-RLS, but to keep the discussion simple, we focus on AF-LMS and ESN. For these AF part of ds-AF-LMS or ds-ESN, we take the publicly available code from \url{http://www.fecgsyn.com}, and follow the suggested parameters accompanying the code.

Lastly, we mention that to the best of our knowledge, less is published about two aECG channels approach (for example, in \cite{Rodrigues2014}, the considered algorithm can be applied to the two channel aECG), and our proposed method focuses on this direction. The main innovation of our approach, compared with other methods, is twofold. First, based on the geometry of the inherited oscillatory structure of the cardiac activity, the diffusion-based manifold learning technique is applied to do the channel section. While other channel selection criteria mainly are based on the power spectral distribution, wave morphology entropy, root mean square error, etc, to find the clearest and most enhanced QRS complexes \cite{DiMariaLiu2014,Ghaffari2015}, our approach is different since we carefully examine the nontrivial underlying geometric structure hosting the cardiac activity by the DM and look for the linear combination that is most like a simple closed curve.
Second, we apply the modern time-frequency analysis technique, the dsSTFT, and the beat tracking algorithms detailed in \cite{Su_Wu:2016} to obtain an accurate R peak locations, and the nonlocal median, to better estimate the maternal ECG morphology and fetal ECG morphology. Compared with other available algorithms, we use more information hidden in the aECG, including decomposing the non-sinusoidal oscillatory pattern from the time-varying frequency, and the low dimensional parametrization of all possible cardiac oscillations. 
We mention that an important advantage of the approach in \cite{Su_Wu:2016} is the ability to separate mECG and fECG with temporal overlap by the nonlocal median. Furthermore, due to its nonlocal nature, it can directly handle a long signal without dividing it into small fragments. 
Notice that unlike the traditional AF-like methods, $\ours$ does not cancel the maternal ECG in one channel by designing a filter from another channel; instead, it directly cancels the maternal ECG in a single linear combination, as is mentioned in Step 1.

\subsection{Materials}

We validate the proposed two-channel algorithm on two publicly available databases of aECG signals. 

\begin{wrapfigure}{R}{0.5\textwidth}
\centering
\includegraphics[width=.5\textwidth]{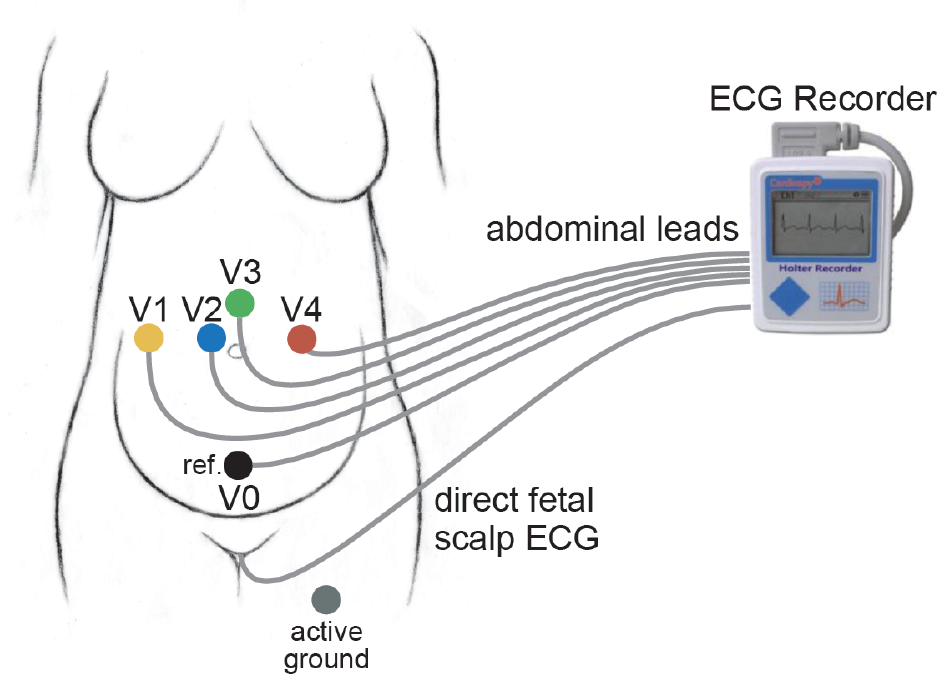}
\caption{The lead placement for the adfecgdb.}
\label{Figure:LeadPlacement}
\end{wrapfigure}

The first database is the PhysioNet non-invasive fECG database ({\em adfecgdb}), where the aECG signals with the annotation provided by experts are publicly available\footnote{\url{https://www.physionet.org/physiobank/database/adfecgdb/}} \cite{Goldberger_Amaral_Glass_Hausdorff_Ivanov_Mark_Mietus_Moody_Peng_Stanley:2000,Kotas2010}.
There are five pregnant women between 38 to 40 weeks of pregnancy in this database. Each has 4 aECG channels and one direct fECG signal recorded from the Komporel system (ITAM Institute, Zabrze, Poland\footnote{\url{http://www.itam.zabrze.pl/developments-english-version-233/665-komporel}}). The four abdominal leads are placed around the navel, a reference lead is placed above the pubic symphysis, and a common mode reference electrode with active-ground signal is placed on the left leg. See Figure \ref{Figure:LeadPlacement} 
for an illustration of the leads placement.
The signal lasts for 5 minutes and is sampled at a fixed rate 1000Hz with the 16bit resolution. The R peak annotation is determined from the direct fECG recorded from the fetal scalp lead.

The second database is the 2013 PhysioNet/Computing in Cardiology Challenge\footnote{\url{https://physionet.org/challenge/2013/#data-sets}}, abbreviated as CinC2013. We focus on the set A composed of 75 recordings for an assessment of our proposed algorithm since it is the only one with the provided the R peak annotation with reference to a direct FECG signal, acquired from a fetal scalp electrode. 
Each recording includes four noninvasive mECG channels that were obtained from multiple sources using a variety of instrumentations with differing frequency response, resolution, and configurations. Although they are from different resources, all recordings are resampled at the sampling rate 1000 Hz and last for 1 minute. There is no publicly available information about where the leads are placed on the maternal abdomen. Note that some recordings come from the adfecgdb database, but no detail is available publicly.  
More details about these two databases can be found on the website. 
We follow the suggestion in \cite{AndreottiRiedl2014} to disregard the recording a54 since it was discarded by the Challenge's organizers, and focus on the remaining 74 recordings.

\subsection{Evaluation metrics}
In the whole analysis, the R peak detection result is evaluated by beat-to-beat comparisons between the detected beats and the provided annotations. We follow the criterion in \cite{Guerrero-Martinez2006} and choose a matching window of 50 ms. Denote $TP$, $FP$, and $FN$ to be true positive rate, false positive rate, and false negative rate,
where $TP$ means correctly detected peaks, $FP$ means nonexistent peaks that were falsely
detected, and $FN$ means existing peaks that were not detected.

We report the sensitivity (SE) and the positive predictive value (PPV) defined as 
\begin{equation}
SE:=\frac{TP}{TP+FN}, \quad PPV=\frac{TP}{TP+FP},
\end{equation} 
and the $F_1$ score, which is the harmonic mean of PPV and SE,
\begin{equation}
F_1:=\frac{2TP}{2TP+FN+FP}.
\end{equation}
We also report the mean absolute error (MAE) of the estimated R peak locations. We follow the suggestion in \cite{Andreotti2016} to report the MAE only on true positive annotations to make the evaluation independent of the detection accuracy. Thus, the MAE is defined as
\begin{equation}
MAE:=\frac{1}{n_{TP}}\sum_{j=1}^{n_{TP}}|r^f_i-\tilde{r}^f_i|,
\end{equation}
where $n_{TP}$ is the number of true positive annotations, and $\tilde{r}^f_i$ and $r^f_i$ are the  temporal location of the $i$-th true positive reference R-peak and temporal location of the $i$-th true positive detected R peak.

For each database, we will report two sets of statistics. First, for each subject, we record the best $F_1$ result among all pairs of available channels, denoted as $F_1(1)$ and report the mean and median of the $F_1(1)$ of all subjects, and the corresponding summary statistics of the MAE, denoted as MAE(1). To see how stable the algorithm is, we also record the median $F_1$ result among all pairs of available channels, called $F_1(0.5)$, and report the mean and median of the $F_1(0.5)$ of all subjects, as well as the corresponding summary statistics of the MAE, denoted as MAE(0.5). Second, to evaluate the lead placement issue, for each pair of available channels, we report the the mean and median of the $F_1$ of all subjects, and the corresponding summary statistics of the MAE. To avoid the boundary effect inevitable in the dsSTFT algorithm due to the window length, the first and last 2 seconds in every recording are not evaluated. 
The notation $a\pm b$ indicates the mean $a$ with the standard deviation $b$.

\subsection{Parameters}
For a fair comparison and the reproducibility purposes, here we summarize the parameters for $\ours$. The parameters are fixed for all signals throughout the paper unless otherwise stated.
For the linear combination of two channels, we fix $\mathcal{I}=\{ -1+k/6\}_{k=1}^{12}$.
The window length $L_{\mathtt{MF}}$ of the median filter for the baseline wandering removal is chosen to be $0.1$ second. 
For the dsSTFT, the beat tracking, and the nonlocal median, the parameters are set to be the same as those reported in \cite{Su_Wu:2016}. 

For the channel selection, we set the lag to $L=7$ for the lag map; we choose the Gaussian kernel and $\alpha=1$ normalization for the DM; we choose $T_{\texttt{CS}}=40$, $n_{\texttt{CS}}=6$ and $L_{\texttt{CS}}=0.1375$ Hz for the adfecgdb database, and $T_{\texttt{CS}}=10$, $n_{\texttt{CS}}=6$ and $L_{\texttt{CS}}=0.25$ Hz for the CinC database. 
We mention that the above parameters are chosen in the ad-hoc fashion without any optimization pursue. Those parameters could be optimized based on the application field and the environment. 

The algorithms are tested on MacBook Air (13-inch, Mid 2013) with Processor 1.3GHz Intel Core i5, Memory 4 GB1600MHz DDR3, Mac OS Sierra (Version 10.12.2), and Matlab R2015b without implementing the parallel computation.

\section{Results}\label{Section:Results}

For the adfecgdb database, 
the direct fECG measurement was lost between 187 and 191 s and between 203 and 211 s in the r10 record, and these two segments were discarded in the evaluation. 
The evaluation results of our proposed algorithm for each combination of two channels out of four available channels of all subjects in the adfecgdb database are shown in Table \ref{tab: adfecgdbF1_CS} for a clear comparison purpose. Except the combination of Channel 2 and Channel 3 in r01 and r08, all the other combinations have the $F_1$ consistently greater than $94$\%. For the MAE, the result is always smaller than 9ms except the combination of Channel 2 and Channel 3 in r08. 
Table \ref{tab:summary_short} shows the comparison of the proposed method with other available algorithms. The $F_1(1)$ and $F_1(0.5)$ of all 6 pairs for each subject are recorded, and the summary statistics of all subjects are shown. It is clear that $\ours$ is consistently better than the other algorithms. 
The average running time is 141.55s for $\ours$,  194.44s for the ds-AF-LMS, 589.83s for the ds-AF-ESN, 10.01s for JADE-ICA, and 10.98s for PCA.

\begin{table}
\centering
\caption{$F_1$ score, mean absolute error (MAE), positive predictive value (PPV), and sensitivity (SE) of all pairs of two channels out of four available channels and all subjects over the whole $5$ minute signals in the adfecgdb database.}
\begin{tabular}{|c|c|cccc|}
\hline
Subject & Channel & $F_1$ (\%) & MAE (msec) & PPV (\%) & SE (\%) \\
\hline
\hline
\multirow{6}{*}{r01} 
& 1 and 2 & 99.45 & 1.35 & 99.22 & 99.69\\
& 1 and 3 & 99.69 & 1.98 & 99.53 & 99.84\\
& 1 and 4 & 99.37 & 2.44 & 99.22 & 99.53\\
& 2 and 3 & 86.94 & 4.44 & 85.87 & 88.03\\
& 2 and 4 & 98.74 & 2.13 & 98.59 & 98.9\\
& 3 and 4 & 99.21 & 2.17 & 99.06 & 99.37\\
\hline
\multirow{6}{*}{r04}
& 1 and 2 & 97.68 & 8.08 & 97.44 & 97.91\\
& 1 and 3 & 97.52 & 8.18 & 97.28 & 97.75\\
& 1 and 4 & 98.72 & 7.42 & 98.56 & 98.88\\
& 2 and 3 & 98.4 & 7.78 & 98.24 & 98.56\\
& 2 and 4 & 98.4 & 8.4 & 98.09 & 98.72\\
& 3 and 4 & 98.72 & 7.42 & 98.56 & 98.88\\
\hline
\multirow{6}{*}{r07}
& 1 and 2 & 98.38 & 8.68 & 98.23 & 98.54\\
& 1 and 3 & 99.03 & 7.47 & 99.03 & 99.03\\
& 1 and 4 & 99.84 & 8.06 & 99.84 & 99.84\\
& 2 and 3 & 99.11 & 8.55 & 99.03 & 99.19\\
& 2 and 4 & 99.27 & 8.59 & 99.19 & 99.35\\
& 3 and 4 & 99.84 & 8.44 & 99.84 & 99.84\\
\hline
\multirow{6}{*}{r08}
& 1 and 2 & 97.6 & 2.18 & 96.78 & 98.44\\
& 1 and 3 & 99.3 & 2.22 & 98.92 & 99.69\\
& 1 and 4 & 99.69 & 1.87 & 99.38 & 100\\
& 2 and 3 & 28.55 & 10.63 & 34.83 & 24.18\\
& 2 and 4 & 97.05 & 2.01 & 96.46 & 97.66\\
& 3 and 4 & 94.36 & 4.87 & 93.43 & 95.32\\
\hline
\multirow{6}{*}{r10}
& 1 and 2 & 98.88 & 2.85 & 98.41 & 99.36\\
& 1 and 3 & 98.88 & 2.85 & 98.41 & 99.36\\
& 1 and 4 & 98.88 & 2.85 & 98.41 & 99.36\\
& 2 and 3 & 98 & 3.37 & 97.46 & 98.55\\
& 2 and 4 & 94.67 & 4.51 & 93.7 & 95.66\\
& 3 and 4 & 94.67 & 4.51 & 93.7 & 95.66\\
\hline
\end{tabular}
\label{tab: adfecgdbF1_CS}
\end{table}

\begin{table}
\centering
\caption{The summary statistics of different methods' performance, including $F_1$ and mean absolute error (MAE), evaluated in the adfecgdb database. The $F_1(1)$ result from the six pairs of two channels is recorded for each subject, and the summary statistics of all subjects is reported in the first ten rows; the $F_1(0.5)$ result from the six pairs of two channels is recorded for each subject, and the summary statistics of all subjects are reported from the 11-th to the 20-th rows. std: standard deviation. $Q_1$: the first quartile. $Q_3$: the third quartile.}
\begin{tabular}{|c||c|ccccc|}
\hline
 &Method & mean & std & $Q_1$ & Median & $Q_3$ \\
\hline
\hline
 & $\ours$ &	99.36 &0.52 & 98.84 &99.69 &99.73  \\ 
\cline{2-7}
 & ds-AF-LMS&	99.55 &	0.74 & 99.44 &99.84 &99.88 \\ 
\cline{2-7}
$F_1(1)$ (\%) & ds-ESN & 99.00 & 1.21 & 98.36 & 99.36 & 99.88  \\
\cline{2-7}
over 6 pairs & JADE-ICA&	39.34 &	34.90 & 17.30 &18.64 &58.85\\ 
\cline{2-7}
& PCA &	50.14 &	41.98 & 18.54 & 22.08 & 95.11\\ 
\hline
  & $\ours$ &	4.44 & 3.05 & 1.96 &2.85 &7.58 \\ 
\cline{2-7}
&ds-AF-LMS& 4.42 & 3.02 & 2.12 & 2.49& 7.64\\ 
\cline{2-7}
MAE(1) (msec)  &ds-ESN& 4.85 & 2.68 & 2.61 & 4.18 &7.62\\ 
\cline{2-7}
over 6 pairs &JADE-ICA&	16.54 & 10.49 & 6.33 &23.41 &24.43\\   
\cline{2-7}
&PCA   &	14.24 & 10.50 & 3.51 &16.32 &24.01\\   
\hline\hline
 & $\ours$ &	98.53 & 0.79 & 98.13 &98.44 &99.22 \\  
\cline{2-7}
  & ds-AF-LMS&	98.46 & 1.69 & 98.01 &98.91 &99.40\\ 
\cline{2-7}
$F_1(0.5)$ (\%) & ds-ESN&	96.66 & 3.99 & 95.03 &98.41 &99.00\\ 
\cline{2-7}
over 6 pairs & JADE-ICA&	21.95 &	7.54 & 16.54 &17.81 &28.11\\ 
\cline{2-7}
& PCA  &	21.41 &	8.40 & 17.55 &17.90 &22.69\\ 
\hline
  & $\ours$ &	4.78 &3.16 & 2.19 & 3.11 & 8.07 \\ 
\cline{2-7}
 &ds-AF-LMS& 5.09 & 2.57 & 3.18 & 3.93 &7.80\\ 
\cline{2-7}
MAE(0.5) (msec) &ds-ESN& 5.64 & 2.30 & 3.74 & 4.94 &7.97\\ 
\cline{2-7}
over 6 pairs &JADE-ICA&	21.58 &	5.64 & 16.48 &24.39 &25.76\\  
\cline{2-7}
&PCA   &	20.52 & 4.94 & 17.74 &19.60 &25.18\\  
\hline
\end{tabular}
\label{tab:summary_short}
\end{table}

For the CinC2013 database,
in Tables \ref{tab:cinc_best} we compare $\ours$ with the other available algorithms in the CinC2013 database. The $F_1(1)$ of all recordings of our method is $92.99\pm 16.0\%$ and the corresponding MAE(1) is $5.38\pm 4.52$ msec, which are both better than the other compared methods. The median $F_1(0.5)$ of all recordings of our method is $85.44 \pm 22.42\%$ and the MAE(0.5) of our method is $6.54 \pm 4.92$ msec, which are both better than the best result determined by other methods. 
It should be noted that the median of $F_1(0.5)$ over 6 pairs of our proposed algorithm is still as high as $96.32$\%, while other methods decline dramatically to less than $60$\%. This result suggests the stability of the proposed method.\footnote{It is suggested in \cite[p.1569]{BeharOsterClifford2014} to remove six more recordings, a33, a38, a47, a52, a71, and a74, in addition to a54, because of some inaccurate reference annotations identified by the visual inspection of authors in \cite{BeharOsterClifford2014}. The $F_1(1)$ of all recordings of our method is $94.80\pm 13.17\%$ and the MAE(1) is $5.04\pm 3.88$ msec, and the $F_1(0.5)$ of all recordings of our method is $87.04 \pm 21.27\%$ and the MAE(0.5) of our method is $6.29 \pm 4.56$ msec.}
The average running time is 20.29s for $\ours$,  27.26s for the ds-AF-LMS, 100.35s for the ds-ESN, 3.29s for JADE-ICA, and 3.20s for PCA.

\begin{table}
\centering
\caption{The summary statistics of different methods' performance, including $F_1$ and mean absolute error (MAE), evaluated in the CinC2013 database. The subject a54 is removed from the datasets.
The $F_1(1)$ result from the six pairs of two channels is recorded for each subject, and the summary statistics of all subjects is reported in the first ten rows; the $F_1(0.5)$ result from the six pairs of two channels is recorded for each subject, and the summary statistics of all subjects are reported from the 11-th to the 20-th rows. std: standard deviation. $Q_1$: the first quartile. $Q_3$: the third quartile.}
\begin{tabular}{|c||c|ccccc|}
\hline
&Method & mean & std & $Q_1$ & Median & $Q_3$ \\
\hline
\hline  
& $\ours$ &	92.99 &16.00& 95.39 &99.21 & 1 \\   
\cline{2-7}
 &ds-AF-LMS&	72.77 & 27.52 & 51.56 &85.50 &98.92\\ 
\cline{2-7}
$F_1(1)$ (\%) &ds-ESN &	72.04 & 27.61 & 50.88 &82.54 &99.20\\ 
\cline{2-7}
over 6 pairs &JADE-ICA&	36.13 &	23.74 & 20.73 & 26.11 &43.81 \\ 
\cline{2-7}
&PCA   &	35.35 &	23.89 & 20.16 &24.00 &37.97 \\ 
\hline
& $\ours$ & 5.38 &4.52 & 1.96 & 4.03 & 7.82 \\ 
\cline{2-7}
 &ds-AF-LMS &	7.06 & 6.13 & 2.93 &5.36 &7.62\\ 
\cline{2-7}
MAE(1) (msec) &ds-ESN &	6.179 & 4.59 & 2.86 &5.54 &7.42\\
\cline{2-7}
over 6 pairs &JADE-ICA&	15.22 &	7.18 & 8.42 &15.82 &21.93\\  
\cline{2-7}
&PCA     &	16.04 &	7.19 & 9.59 &18.00 &21.75\\  
\hline\hline
& $\ours$ &	85.43 &22.42& 83.27 &96.32 &99.57 \\ 
\cline{2-7}
 &ds-AF-LMS&	56.34 & 30.51 & 25.69 &51.55 &90.38\\ 
\cline{2-7}
$F_1(0.5)$ (\%) &ds-ESN &	58.22 & 30.85 & 36.69 &54.55 &89.54\\ 
\cline{2-7}
over 6 pairs &JADE-ICA&	26.67 &	20.52 & 16.88 &19.51 &24.74\\ 
\cline{2-7}
&PCA    &	26.59 &	21.31 & 15.96 & 19.24 &25.19\\ 
\hline
& $\ours$ &	6.54 & 4.92 & 2.55 &5.70 &5.53 \\ 
\cline{2-7}
 &ds-AF-LMS&	11.63 & 8.0283 &  5.70 &8.50 &18.66 \\  
\cline{2-7}
MAE(0.5) (msec) &ds-ESN &	9.85 & 6.57 &  4.67 &7.88 &13.96 \\ 
\cline{2-7}
over 6 pairs &JADE-ICA&	20.44 &	6.46 & 16.80 & 22.17 &24.97 \\ 
\cline{2-7}
& PCA    &	20.59 &	7.07 & 15.64 & 22.61 &25.24 \\ 
\hline
\end{tabular}
\label{tab:cinc_best}
\end{table}

To further evaluate the influence of the lead placement, or to answer if we could design the best lead placement scheme for the proposed two-channel algorithm, we report the summary statistics of all pairs of two channels for the adfecgdb database in Table \ref{tab: combination_short} and the CinC2013 database in Table \ref{tab: combination_cinc}. 
It is interesting to see that for the adfecgdb database, except for the combination of channel 2 and channel 3, the mean $F_1$ accuracy is great than $97$\%. The outlier of the combination of channel 2 and channel 3 comes from the fact that the fECG is strong in case r08, which confuses the channel selection step. As a result, $\ours$ extracts the maternal ECG as the fECG, which leads to a wrong fECG estimation.\footnote{If we are allowed to use the physiological information that both the fetus and the mother are healthy so that the fetal IHR is on average higher than maternal IHR, then we could correct this confusion by swapping the fetal IHR and maternal IHR. This leads to the mean $F_1$ of the combination of channel 2 and channel 3 $93.44$\% with the standard deviation $6.99$\% and the mean MAE $5.57$ ms with the standard deviation $2.41$ ms, and the results of other combinations unchanged.} While determining the role of each component is a common issue for the fetal-maternal ECG separation algorithms and commonly we need more information to handle it, we leave this open problem for the future work. 

Compared with the result of the adfecgdb database, the performance of $\ours$ in the CinC2013 database is not uniform cross different combinations of channels. Note that the lead placement scheme is unknown for the CinC2013 database, so it is not possible to conclude which pair of channels is the best. However, if we assume that the lead placement scheme for all recordings in the CinC2013 database is the same as the lead placement scheme shown in Figure \ref{Figure:LeadPlacement}, then the CinC2013 database results suggest that the best combination is channel 1 and channel 4; the $F_1$ has the mean of $87.93$\% with the standard deviation $22.64$\%, and the median $97.60$\% with the interquartile range $6.92$\%; the MAE has the mean of $6.21$ ms with the standard deviation $6.03$ ms, and the median $4.34$ ms with the interquartile range $5.62$ ms.\footnote{If we remove a33, a38, a47, a52, a54, a71, and a74 from the CinC2013 database \cite{BeharOsterClifford2014}, for the combination of channel 1 and channel 4, the $F_1$ has the mean $89.81$\% with the standard deviation $20.84$\%, and the median becomes $98.41$\% with the interquartile range $5.10$\%; the MAE has the mean of $5.74$ ms with the standard deviation $5.33$ ms, and the median $4.20$ ms with the interquartile range $5.48$ ms.} 
Another finding deserves a discussion is that unlike the adfecgdb database, we can see the discrepancy between the best $F_1$ out of the 6 pairs reported in Table \ref{tab:cinc_best} and the average $F_1$ of each pair reported in Table \ref{tab: combination_cinc}. This might suggest that the lead system applied in the CinC2013 database is heterogenous across the recordings.

For the adfecgdb database, our result is overall compatible with, or better than, the state-of-art result reported in the field. For example, if we choose the pair of channel 1 and channel 2, our result is better than the best channel result based on the continuous wavelet transform based single-channel algorithm \cite[Table 5]{CastilloMorales2013}. However, it is not a fair comparison since the algorithm used in \cite{CastilloMorales2013} is a single-channel algorithm. On the other hand, if we compare with the methods based on ICA on four channels \cite[Table 1]{Poian2015}, our result is compatible. The MAE, which is less reported in the literature, is as small as $10$ msec, which indicates the potential of applying the $\ours$ to do the fetal heart rate variability (HRV) analysis.

For the CinC2013 database, our result is compatible, or better than, the reported results. At the first glance, it is not the case, since by the ICA-based algorithms \cite{BeharOsterClifford2014,AndreottiRiedl2014}, the accuracy could be as high as have the mean $F_1=96$\%, under the same setup that a detected R-peak was labelled as TP if within $50$ ms of a reference R-peak. However, we mention that unlike $\ours$, these algorithms are ICA-based and four channels are simultaneously used. Specifically, in \cite[Table 3]{BeharOsterClifford2014}, among different combinations of different algorithms, the algorithm FUSE-SMOOTH achieved the best result -- the mean $F_1$ over all recordings is $96$\%, after removing a33, a38, a47, a52, a54, a71, and a74; in \cite[Table 1]{AndreottiRiedl2014}, the augmentation, the ICA, the template adaptation or extended Kalman filter, and other techniques are applied, and the result with the mean $F_1=97.3$\% over all recordings with the standard deviation $0.108$ is reported based on the template adaptation, after removing a54. 
Our proposed algorithm, on the other hand, outperforms the algorithm based on four channels and the PCA, for example, \cite{DiMariaLiu2014}. In \cite[Section 3.2]{DiMariaLiu2014}, the accuracy of the proposed algorithm in detecting the fetal heart beats gives the mean $F_1=89.8$\% over all recordings, under the setup that a detected R-peak was labelled as TP if within $100$ ms of a reference R-peak and removing 9 recordings, including a29, a38, a54, a56, a33, a47, a52, a71, and a74. Another novel method based on the channel selection over 4 channels followed by the sequential total variation denoising \cite[Table 5]{LeeLee2016} leads to the accuracy with $F_1=89.9$\% and the MAE$=9.3$ ms\footnote{In a private communication, the authors confirmed that this $F_1$ is the ``overall $F_1$'', which is evaluated by collecting all beats from all recordings, and evaluate the $F_1$ on all collected beats. If we follow the same procedure and remove a33, a38, a47, a52, a54, a71, and a74, the overall $F_1$ of $\ours$ for the combination of channel 1 and channel 4 is $89.77$\% and the MAE is $4.91$ ms.} under the setup that a detected R-peak was labelled as TP if within $50$ ms of a reference R-peak and removing a33, a38, a47, a52, a54, a71, and a74.
We emphasize that while our algorithm does not outperform some of the above-mentioned algorithms, based on two channels, $\ours$ leads to the MAE as small as $6.21$ ms in channel 1 and channel 4 combination in the CinC2013 database, which again indicates the potential of applying the $\ours$ to do the fetal HRV analysis.

\begin{table}
\centering
\caption{The summary statistics of $\ours$, including $F_1$ and mean absolute error (MAE), for six pairs of four available channels in the adfecgdb database. std: standard deviation. $Q_1$: the first quartile. $Q_3$: the third quartile.}
\begin{tabular}{|c||c|ccccc|}
\hline
&Channels & mean  & std & $Q_1$ & Median & $Q_3$ \\
\hline
\hline
\multirow{6}{*}{$F_1$ (\%)}  
&  1 and 2  &	98.40 & 0.79 &97.66 & 98.38 &99.02 \\ 
\cline{2-7}
& 1 and 3& 98.88 & 0.82 &98.54 & 99.03 &99.40 \\ 
\cline{2-7}
&1 and 4 &	99.30 &	0.49 & 98.84 &99.37 &99.73\\ 
\cline{2-7}
&2 and 3& 82.20 & 30.41 & 72.34 &98.00 &98.58\\ 
\cline{2-7}
&2 and 4& 	97.63 &	1.85 & 96.46 &98.4 &98.88\\ 
\cline{2-7}
& 3 and 4 & 97.36 & 2.63 & 94.59 &98.72 &99.37\\ 
\hline
\hline
\multirow{6}{*}{MAE (msec)}  &  1 and 2  &	4.63 & 3.47 & 1.98 & 2.85 &8.23 \\ 
\cline{2-7}
& 1 and 3&   4.54 & 3.02 & 2.16 & 2.85 &7.64 \\ 
\cline{2-7}
&1 and 4 &	4.53 & 2.96 & 2.30 & 2.85 &7.58\\ 
\cline{2-7}
&2 and 3& 	6.95 & 3.00 & 4.17 & 7.78 &9.07\\ 
\cline{2-7}
&2 and 4& 	5.13 & 3.23 & 2.10 & 4.51 &8.45\\ 
\cline{2-7}
& 3 and 4 & 	5.48 & 2.49 & 3.93 & 4.87 &7.67\\ 
\hline
\end{tabular}
\label{tab: combination_short}
\end{table}

\begin{table}
\centering
\caption{The summary statistics of $\ours$, including $F_1$ and mean absolute error (MAE), for six pairs of four available channels in the CinC2013 database. The subject a54 is removed from the datasets. std: standard deviation. $Q_1$: the first quartile. $Q_3$: the third quartile.}
\begin{tabular}{|c||c|ccccc|}
\hline
&Channels & Mean & std & $Q_1$ & Median & $Q_3$ \\
\hline
\hline
\multirow{6}{*}{$F_1$ (\%)}  & 1 and 2  &	81.69 &25.82& 72.60 &95.62 &99.36 \\ 
\cline{2-7}
& 1 and 3& 82.93 & 26.28 &83.27 & 96.24 &99.36 \\ 
\cline{2-7}
& 1 and 4 & 87.93& 22.64 & 93.08 &97.60 &1\\ 
\cline{2-7}
& 2 and 3&	74.40 &30.63 & 36.10 &93.33 &99.67\\ 
\cline{2-7}
& 2 and 4&	81.50 &	26.64 & 66.95 &96.51 &99.36\\ 
\cline{2-7}
& 3 and 4 & 79.83 & 28.49 & 58.78 &96.96 &99.67\\ 
\hline
\hline
\multirow{6}{*}{MAE (msec)}  & 1 and 2  &	7.72 &7.03 & 2.53 &5.04 &9.32 \\ 
\cline{2-7}
& 1 and 3 & 7.83 & 7.45 & 2.42 & 6.08 & 8.74 \\ 
\cline{2-7}
& 1 and 4 & 6.21 &  6.03 & 2.04 &4.34 &7.66\\ 
\cline{2-7}
& 2 and 3&	9.44 &6.88 & 4.12 &8.05 &12.97\\ 
\cline{2-7}
& 2 and 4&	7.93 &6.62 & 3.61 &6.28 &9.67\\ 
\cline{2-7}
& 3 and 4 & 7.85 & 6.85 & 2.31 & 5.92 & 9.97\\ 
\hline
\end{tabular}
\label{tab: combination_cinc}
\end{table}

\section{Discussion}\label{Section:Discussions}

The encouraging results of $\ours$ indicate the possibility to design a ``two-lead system'' for the noninvasive, and long term fECG monitoring purpose. 
As discussed above, theoretically, the chance is low that the fetal cardiac axis orientation would be so much orthogonal to the 2-dim affine subspace spanned by the two leads that no fECG shape can be reconstructed. This is a big advantage compared with the single-lead system, as the chance that the fetal cardiac axis orientation is orthogonal to the 1-dim affine subspace spanned by the single lead is much higher. Thus, while there have been several successful algorithms for the one aECG channel, like \cite{CastilloMorales2013,Behar2014,Su_Wu:2016} and the citations inside, if the recorded one channel signal does not have fECG information, there is nothing the algorithm can do.
From the practical viewpoint, since only two leads are needed, the corresponding hardware could be lighter and more deployable than the currently available four-lead or multiple-lead systems. 
While it is certainly possible to generalize our algorithm to a three-lead or four-lead system (and the algorithm can be changed directly according to the setup), to have a better balance between the prediction accuracy, the hardware design, and practical purposes, we focus on the two-lead system in our research.

Despite of the above-mentioned benefits, there are several challenges we need to solve until this possible system is clinically usable.
As is shown above, the performance of $\ours$ depends on how the two leads are put on the abdomen. The fECG situation is clearly different from the adult ECG system, like the widely applied 12 lead ECG system. Since fetus does move and rotate inside the uterus, the uterus differs from female to female, and the maternal body profile varies, we may not expect to have a two-lead system universal for all women. Therefore, for the practical purpose, particularly for the long term monitoring purpose and the future digital health, like the wearable biosensors \cite{Li_Dunn_Salins:2017}, it is important to ask if we could adaptively find the best lead placement scheme for different females. For the practical purpose, due to the inevitable non-stationary noise of different types, like the motion artifact and uterine contraction, an automatic system providing a SQI to alarm/warn the low quality of the lead system, and hence improve the overall fECG extraction quality, is urgently needed. We leave this important engineering problem to the future work.
Another interesting question naturally raises from the current work is if we could generalize the current algorithm to study the twin dataset. Theoretically it is possible, if we take the fact that geometrically the twin will locate in different positions. We would expect to study this problem when the dataset is available.

From the algorithmic viewpoint, there are several directions we could improve the proposed two-channel fECG algorithm. The main ingredient in $\ours$ is the diffusion geometry. Since we have more than one aECG channel, we could consider modern diffusion-based manifold learning technique to extract information common in two channels, like the alternating diffusion \cite{lederman2015alternating,2016arXiv161108472P,talmon2016latent}. The non-stationary nature of the fECG signal, which often presents itself as a time-varying frequency, might jeopardize the diffusion-based approach. We could consider to entangle the nontrivial time-varying frequency nature of the signal by further applying the modern nonlinear-type time-frequency analysis technique, like the synchrosqueezing transform or concentration of frequency and time (see \cite{Daubechies_Wang_Wu:2016} and the citations inside).

Another important algorithmic question left unanswered in this paper is how to improve the nonlocal median algorithm so that the reconstructed fECG could provide more accurate electrophysiological information about the heart, for example, the ECG morphology like the Q wave and ST-segment section information \cite{Amer_Hellsten_Noren:2001}. The main difficulty encountered in this problem is the lack of the ``ground truth'', and a careful design of the clinical trial to acquire a reliable ground truth for the morphological study of the fetal cardiac activity is needed.
As important as this clinical information could be, we will focus on it as an independent research and report the result in the future work. 

Last but not the least, the databases we tested are small and not specifically designed for our purpose. We thus need a large scale and well designed prospective study to confirm the result.

\section{Conclusion}\label{Section:Conclusion}

A novel two-channel fetal-maternal ECG signal separation algorithm, $\ours$, is proposed. The potential of the proposed algorithm is supported by the positive validation results on two publicly available databases. The algorithm is both computationally efficient and is supported by the underlying rigorous mathematical model and theory. Its clinical applicability will be evaluated in the future work.

\section{Acknowledgement}
Hau-tieng Wu's research is partially supported by Sloan Research Fellow FR-2015-65363. The authors acknowledge the help of Jan Hamanishi to prepare the illustration in Figure \ref{Figure:LeadPlacement}.

\bibliographystyle{plain}
\bibliography{fECG,fECGCepstrum}

\appendix

\renewcommand{\thefigure}{SI.\arabic{figure}}
\renewcommand{\thetable}{SI.\arabic{table}}
\renewcommand{\theequation}{S.\arabic{equation}}
\renewcommand{\thesection}{SI.\arabic{section}}
\renewcommand{\thesection}{SI.\arabic{section}}

\section{Theoretical background for the channel selection}\label{Section:TheoreticalBackground}

We describe the background material needed for the proposed two-channel fECG algorithm. The algorithm is composed of three essential components. 
The first component is estimating the maternal cardiac activity in the aECG (maECG) by applying the dsSTFT \cite{Lin_Li_Wu:2016}, the beat tracking and the nonlocal median \cite{Su_Wu:2016}, and get a rough fECG from any given linear combination of the two provided aECG signals.
The second component is the channel selection by applying the lag map \cite{Takens:1981,Richter1998,Kotas2010} and diffusion map (DM) \cite{Coifman_Lafon:2006} for the sake of determining the best linear combination of the two channels, which lead to the optimal rough fECG.
The third component is getting the fECG and fetal R peaks information from the optimal rough fECG by again the dsSTFT and the beat tracking.

\subsection{Linear combination}
The main motivation behind the algorithm is motivated by the physiological knowledge of the ECG signal that among all linear combinations of two channels, with a high probability we could find a combination that is optimal for the fetal ECG extraction.

Before describing the linear combination idea, recall the well-know {\em vectocardiogram} (VCG) and its relationship with the ECG signals.
It has been well known that the ECG signal, denoted as a continuous time series $E:[0,T]\to \mathbb{R}$, where $T>0$ is the observation time, is the projection of the representative dipole current of the electrophysiological cardiac activity on a predesigned direction \cite{Keener:1998}. Denote the dipole current as a three dimensional continuous time series $d:[0,T]\to\mathbb{R}^3$. If we could record $d(t)$, it is called the VCG signal. 
Physiologically, for a normal subject, $d(t)$ is oscillatory with the period $\tau>0$, which is about $1$ second, in the sense that $d(t)\sim d(t+\tau)$ for all $t\in[0,T-\tau]$. Suppose $t_l$, $l=1,\ldots,m$, where $m$ is the number of cardiac cycles over the period $[0,T]$, is the timestamp corresponding to the maximal amplitude point of the $l$-th cardiac cycle.  
We call the vector 
\begin{align}
c=\frac{1}{m}\sum_{l=1}^md(t_l)
\end{align}
the cardiac axis. 
For a given ECG signal, there is an associated projection direction $v\in \mathbb{R}^3$ so that $E$ is the projection of $d(t)$ on $v$; that is, $E(t)=v^Td(t)$. It has been well known that depending on $v$, we could acquire different aspects of the cardiac information. 
We mention that in general, $v$ changes according to time due to the cardiac axis deviation caused by the respiratory activity and other physical movements. To simplify the discussion, we do not take these facts into account. 

Denote $d_m$ to be the mother's VCG and $d_f$ to be the fetus' VCG. Denote $c_m$ to be the mother's cardiac axis and $c_f$ to be the fetus' cardiac axis. Fix two abdominal lead placements and record two aECG signals, denoted as $x_1$ and $x_2$. Denote $v_{m,i}\in \mathbb{R}^3$ and $v_{f,i}\in \mathbb{R}^3$ to be the projection directions of the mother's VCG and fetus' VCG corresponding to $x_i$, where $i=1,2$. Obviously, we have $x_i=v_{m,i}^Td_m+v_{f,i}^Td_f$, where $i=1,2$, and it is possible that the fetal cardiac activity is relatively weak in both $x_1$ and $x_2$. To resolve this problem, we consider the following linear combination scheme. Take a linear combination of $x_1$ and $x_2$ by 
\begin{align}
x_\theta&=\theta x_1+\sqrt{1-\theta^2}x_2\\
&=\theta v_{m,1}^Td_m+\theta v_{f,1}^Td_f +\sqrt{1-\theta^2}v_{m,2}^Td_m+\sqrt{1-\theta^2}v_{f,2}^Td_f\nonumber\\
&=[\theta v_{m,1}^T+\sqrt{1-\theta^2}v_{m,2}^T]d_m+[\theta v_{f,1}^T+\sqrt{1-\theta^2}v_{f,2}^T]d_f\nonumber, 
\end{align} 
where $\theta\in (-1,1]$. 
If these two abdominal leads are placed on two different locations, we know $v_{f,1}\neq v_{f,2}$ and $v_{m,1}\neq v_{m,2}$, and hence the set 
\begin{equation}
A:=\{\theta v_{f,1}-\sqrt{1-\theta^2}v_{f,2}\}_{\theta\in (-1,1]}
\end{equation}
contains all linear combinations of $v_{f,1}$ and $v_{f,2}$ if we do not distinguish $\theta v_{f,1}-\sqrt{1-\theta^2}v_{f,2}$ and $-\theta v_{f,1}+\sqrt{1-\theta^2}v_{f,2}$. Note that the set $\{(\theta,\sqrt{1-\theta^2})\}_{\theta\in (-1,1]}$ is the 1-dimensional real projective space (identifying antipodal points of the unit circle, $S^1:=\{x\in \mathbb{R}^2| \|x\|=1\}$, embedded in $\mathbb{R}^2$), and it topologically equivalent to the unit circle $S^1$, which is an one dimensional manifold. Based on the above-mentioned relationship between ECG and VCG, although the fetus could rotate inside the uterus, we know that unless the cardiac axis $c_f$ of the fetal cardiac activity is perpendicular or almost perpendicular to both $v_{f,1}$ and $v_{f,2}$, we could find an $\theta$ so that $x_\theta$ contains a strong fetal cardiac activity. 
Since the chance that the fetal cardiac axis is perpendicular to the 2-dim affine subspace corresponding to the two chosen abdominal leads is low, we could thus conclude that the chance that we could obtain a good signal with strong fECG via the linear combination scheme is high.

\subsection{Lag map}

The lag map is a well-known method widely applied to study a given time series, and its theoretical foundation has been well established in \cite{Takens:1981,Stark_Broomhead_Davies_Huke:1997,Stark_Broomhead_Davies_Huke:2003}. In brief, it allows us to reconstruct the structure underlying the time series. 
For a given time series $\boldsymbol{f}$ of length $N\in\mathbb{N}$, the lag map is a mapping from $\boldsymbol{f}$ to a set of $L$-dim points, where $L$ is chosen by the user, via
\begin{equation}
\Psi_{\boldsymbol{f},L}:i\mapsto (\boldsymbol{f}(i),\ldots,\boldsymbol{f}(i+L))^T\in\mathbb{R}^{L+1},
\end{equation}
where $i=1,\ldots, N-L$ and the superscript $T$ means taking the transpose. 
The map $\Psi_{\boldsymbol{f},L}$ is called the {\it $L$-step lag map}. It has been shown in \cite{Takens:1981} that if $\boldsymbol{f}$ is an observation of a dynamical process whose trajectory is supported on a $d$ dimensional manifold and $L$ is large enough, then under some weak mathematical conditions, $\Psi_{\boldsymbol{f},L}$ could recover the manifold up to a diffeomorphism. Since the cardiac activity is periodic, the corresponding ``underlying manifold'' is a one-dimensional circle representing the cardiac dynamics that is diffeomorphic to the unit circle $S^1$, and the lag map of the cardiac activity time series leads to a point cloud supported on another one-dimensional simple closed curve. 

The above-mentioned important property of the lag map allows up to examine the quality of the reconstructed fECG. 
If $\boldsymbol{f}\in\mathbb{R}^N$ is the true fECG signal, or a good estimation of the fECG signal, we obtain an one-dimensional simple closed curve by the point cloud $\mathcal{X}_{\boldsymbol{f},L}:=\{\Psi_{\boldsymbol{f},L}(i)\}_{i=1}^{N-L}\in \mathbb{R}^{L+1}$. 
On the other hand, if the tempted fECG estimator $\boldsymbol{f}\in\mathbb{R}^N$ fails to be a good estimator of the fECG signal, the point cloud $\mathcal{X}_{\boldsymbol{f},L}$ might be away from any one-dimensional simple closed curve. 
Another important fact is that when $\boldsymbol{f}$ is the fECG signal, the point cloud $\mathcal{X}_{\boldsymbol{f},L}$ is in general non-uniformly sampled from the one-dimensional circle. This fact comes from the diffeomorphic relationship between the reconstructed simple closed curve and the underlying simple closed curve via the lag map.

\subsection{Graph Laplacian and diffusion map}
To take this important fact into account to examine the quality of the reconstructed fECG via the $L$-step lag map, we apply the graph Laplacian (GL), which is the building block of several dimension reduction algorithms, like the diffusion map (DM) \cite{Coifman_Lafon:2006,Singer_Wu:2016}. For a general introduction of GL and DM, we refer readers to \cite{Coifman_Lafon:2006,ElKaroui_Wu:2016,Singer_Wu:2016}. Here we only provide the necessary steps for our purpose. 
Fix the given embedded point cloud $\mathcal{X}_{\boldsymbol{f},L}$. Build a complete affinity graph $G=(V,E,\omega)$ with vertices $V=\mathcal{X}_{\boldsymbol{f},L}$ by viewing any pairs of points in $\mathcal{X}_{\boldsymbol{f},L}$ as edges; that is $E=\{(\Psi_{\boldsymbol{f},L}(i),\Psi_{\boldsymbol{f},L}(j))|\,i\neq j\}$. The affinity function $\omega:E\to\mathbb{R}^+$ is then defined by
\begin{equation}
\label{W}
\omega(\Psi_{\boldsymbol{f},L}(i),\Psi_{\boldsymbol{f},L}(j))=\exp\Big\{-\frac{\|\Psi_{\boldsymbol{f},L}(i)-\Psi_{\boldsymbol{f},L}(j)\|^2}{\epsilon}\Big\},
\end{equation}
for $i,j=1,\ldots,N-L$, $i\neq j$, and $\epsilon>0$ is the kernel bandwidth chosen by the user.
Here, the affinity between $\Psi_{\boldsymbol{f},L}(i)$ and $\Psi_{\boldsymbol{f},L}(j)$ is reversely proportional to the distance between $\Psi_{\boldsymbol{f},L}(i)$ and $\Psi_{\boldsymbol{f},L}(j)$. Note that while we could choose a more general kernel, here we focus on the Gaussian kernel to simplify the discussion. We mention that in practice the Gaussian kernel performs well and the dependence on the chosen kernel is marginal. In general, the point cloud might not be uniformly sampled from the geometric object we have interest, and the nonuniform sampling effect might generate a negative impact on the upcoming analysis. To resolve this issue, the $\alpha$-normalization technique is introduced in \cite{Coifman_Lafon:2006}. Take $0\leq \alpha\leq 1$, we could define an $\alpha$-normalized affinity function defined on $E$, denoted as $\omega^{(\alpha)}$, by
\begin{equation}
\omega^{(\alpha)}(\Psi_{\boldsymbol{f},L}(i),\Psi_{\boldsymbol{f},L}(j))=\frac{\omega(\Psi_{\boldsymbol{f},L}(i),\Psi_{\boldsymbol{f},L}(j))}{d_i^\alpha d_j^\alpha},
\end{equation} 
where $d$ is the degree function defined on the vertex set as
\begin{equation}
d_{i}=\sum^{N-L}_{j=1}\omega^{(\alpha)}(\Psi_{\boldsymbol{f},L}(i),\Psi_{\boldsymbol{f},L}(j)),
\end{equation}
for $i=1,\ldots,N-L$.
As is shown in \cite{Coifman_Lafon:2006,Singer_Wu:2016}, when $\alpha=1$, this $\alpha$-normalized affinity could effectively alleviate the impacts introduced by the nonuniform sampling. In our fECG application, as discussed above, $\mathcal{X}_{\boldsymbol{f},L}$ is in general non-uniformly sampled from the one-dimensional simple closed curve, so we apply this $\alpha$-normalization technique.

We are now ready to define the GL. Define an $\alpha$-normalized affinity matrix $W^{(\alpha)}\in \mathbb{R}^{(N-L) \times (N-L)}$ by
\begin{equation}
W^{(\alpha)}_{ij}:=\omega^{(\alpha)}(\Psi_{\boldsymbol{f},L}(i),\Psi_{\boldsymbol{f},L}(j)),
\end{equation}
for $i,j=1,\ldots,N-L$,
define a diagonal $\alpha$-normalized degree matrix $D^{(\alpha)}\in \mathbb{R}^{(N-L) \times (N-L)}$ by
\begin{equation}
D^{(\alpha)}_{ii}=\sum^{N-L}_{j=1}W_{ij},
\end{equation}
for $i=1,\ldots,N-L$.
and the {\it $\alpha$-normalized graph Laplacian} is then defined by
\begin{equation}
L^{(\alpha)}:=I-{D^{(\alpha)}}^{-1} W^{(\alpha)}.
\end{equation}
Since $L$ is similar to the symmetric matrix $I- {D^{(\alpha)}}^{-1/2} W^{(\alpha)} {D^{(\alpha)}}^{-1/2}$, it has a complete set of right eigenvectors $\phi_1,\ldots,\phi_{N-L}$ with corresponding eigenvalues $0=\lambda_1<\lambda_2\leq\dots\leq\lambda_{N-L}\leq1$. Note that $\phi_1=(1,1\dots,1)^T$ since ${D^{(\alpha)}}^{-1} W^{(\alpha)}$ is a transition matrix defined on the graph $G$. It has been shown in \cite{Coifman_Lafon:2006,Singer_Wu:2016} that if $\mathcal{X}$ is sampled from a low dimensional Riemannian manifold, when $\alpha=1$ and $N\to \infty$, asymptotically the eigenvectors $\phi_i$ converges pointwisely and spectrally to the $i$-th eigenfunction of the Laplace-Beltrami operator of the Riemannian manifold.
In general, this allows us to reconstruct the manifold by applying the diffusion geometry and the spectral embedding theory, which is commonly known as the DM algorithm \cite{Coifman_Lafon:2006}. The robustness of the GL and DM has been studied in \cite{ElKaroui_Wu:2016}. 

In our problem, due to the periodic oscillation intrinsic to the fECG we have interest, the $\alpha$-normalized graph Laplacian associated with $\mathcal{X}_{\boldsymbol{f},L}$ gives us the Laplace-Beltrami operator over a simple closed curve. It follows that asymptotically, the first two non-trivial eigenvectors are the sine and cosine functions.
We could thus take this fact into account and design the signal quality index for the channel selection purpose.

\end{document}